\documentclass[10pt,showpacs,preprintnumbers,amsmath,amssymb]{revtex4}
\usepackage{units}
\usepackage{bbold,euler,newcent}
\usepackage{graphicx}
\usepackage{dcolumn}
\usepackage{bm,bbm,mathtools,esvect}
\usepackage{slashed}

\begin{document}

\title{\LARGE\sc Quaternionic elastic scattering \vspace{5mm}}

\author{\tt\large SERGIO GIARDINO} 
\email{sergio.giardino@ufrgs.br}
\affiliation{\vspace{3mm} Departamento de Matem\'atica Pura e Aplicada, Universidade Federal do Rio Grande do Sul (UFRGS)\\
Avenida Bento Gon\c calves 9500, Caixa Postal 15080, 91501-970  Porto Alegre, RS, Brazil}

\begin{abstract}
\noindent We study the elastic scattering of quantum particles based on a real Hilbert space approach to quaternionic quantum mechanics 
($\mathbbm H$QM)  and derive expression for the wave function, the phase shifts, as well as the optical theorem for the case of a hard sphere scattering potential. The strong agreement between these new quaternionic results and the corresponding results in complex quantum mechanics reinforce the validity of the $\mathbbm H$QM generalization of ordinary complex quantum mechanics ($\mathbbm C$QM)
\end{abstract}

\maketitle
\section{\;\sc Introduction\label{I}}
The mathematical framework of quantum mechanics is based on the complex number field and complex wave functions in a complex Hilbert space.
However, one may ask what would happen to quantum mechanics after replacing the complexes by another number field. Would this theory either be a generalization, a different theory or simply a nonsense speculation? In fact, this question was firstly asked by
John von Neumann and Garrett Birkhoff \cite{Birkhoff:2017kpl}, while  
Ernst Stueckelberg later achieved the development of real quantum mechanics ($\mathbbm{R}$QM)
\cite{Stueckelberg:1960rsi,Stueckelberg:1961rsi,Stueckelberg:1961psg,Stueckelberg:1962fra}, which is considered equivalent to $\mathbbm{C}$QM.
A quaternionic quantum theory was later developed using anti-hermitian Hamiltonian operators, and is summarized in a comprehensive book by Stephen Adler \cite{Adler:1995qqm}. However, the anti-hermitian proposal presents several shortcomings, most importantly the ill-defined classical limit. 
Furthermore, anti-hermitian $\mathbbm{H}$QM solutions are few, involved, and difficult to interpret. By way of example, we quote several results of this kind
\cite{Davies:1989zza,Davies:1992oqq,Ducati:2001qo,Nishi:2002qd,DeLeo:2005bs,Madureira:2006qps,Ducati:2007wp,Davies:1990pm,DeLeo:2013xfa,DeLeo:2015hza,Giardino:2015iia,Sobhani:2016qdp,
Procopio:2017vwa,Sobhani:2017yee,Hassanabadi:2017wrt,Hassanabadi:2017jiz,Sobhani:2017nfa,Bolokhov:2017ndw,DeLeo:2019bcw,Muraleetharan:2014qma,Sabadini:2017qma}.
We also point out the existence of quaternic applications  in quantum mechanics that cannot be considered $\mathbbm{H}$QM, such as \cite{Arbab:2010kr,Kober:2015bkv,Brody:2011mg,Tabeu:2019cqw,Chanyal:2017rqw,Chanyal:2017xqf,Chanyal:2019gdi,Chanyal:2019zse,Cahay:2019bqp,Cahay:2019pse,Marques-Bonham:2020fmw}.  

More recently, a novel approach to $\mathbbm{H}$QM eliminated the anti-hermiticity requirement for the Hamiltonian \cite{Giardino:2018lem} using a real Hilbert space \cite{Giardino:2018rhs}. This approach enabled us to obtain a well-defined classical limit  \cite{Giardino:2018lem} and a proof of the Virial theorem \cite{Giardino:2019xwm}. Furthermore, novel explicit solutions of $\mathbbm H$QM were turned possible using this approach, such as the Aharonov-Bohm effect \cite{Giardino:2016abe}, the free particle \cite{Giardino:2017yke,Giardino:2017nqs}, the square potential solutions \cite{Giardino:2020cee}, the quantum Lorentz law \cite{Giardino:2019xwm} and the scattering from a delta function \cite{Hasan:2020ekd}.  We can confidently state that the real Hilbert space $\mathbbm{H}$QM is physically consistent in a higher level than the anti-Hermitian $\mathbbm{H}$QM, and more powerful in terms of their results.

In this article we show that the quantum elastic scattering  can be obtained in terms of  the real Hilbert space $\mathbbm H$QM, a result  that was never obtained using the anti-hermitian theory. 
The scattering in $\mathbbm{C}$QM is well established and currently described in several textbooks, while
the quaternionic scattering contains a wide and unexplored territory that  was initially supposed to give equivalent results when compared to the complex theory \cite{Adler:1987gd,Adler:1995qqm}. Recent results contradict this version, and quaternionic effects were ascertained in a variety of systems, such as non-relativistic anti-hermitian
\cite{Horwitz:1993by,Horwitz:1993bx,DeLeo:2019bcw}, non-relativistic non-Hermitian \cite{Hasan:2020ekd} and relativistic anti-hermitian \cite{Hassanabadi:2017wrt,Sobhani:2017yee}. Supporting these results, in the present paper we show that the quaternionic scattering of spherically
symmetric potentials contradicts  the hypothesis of an asymptotically complex $\mathbbm{H}$QM because the quaternionic wave function admits a cross section quintessentially different from that found in the complex case. A quaternionic optical
theorem is also obtained, and  it has a natural continuity in the research of the scattering matrix and the stationary collision theory. As a concluding remark, our results support the view of the real Hilbert space $\mathbbm H$QM as physically different from $\mathbbm C$QM, and not simply a mathematical  generalization of quantum mechanics. The precise definition of such a theory is an exciting direction for future research.


\section{\;\sc Wave function\label{W}}

Before considering the proposals to $\mathbbm{H}$QM, we notice that a comprehensive introduction to quaternions ($\mathbbm H$) is beyond the scope of this article, and can be found elsewhere \cite{Ward:1997qcn,Rocha:2013qtt,Morais:2014qch}. We simply remark that quaternions are non-commutative hyper-complex numbers that encompass three imaginary units, $i,\,j$ and $k$, and that a general quaternion  $q$ is such that
\begin{equation}\label{i1}
 q=x_0 + x_1 i + x_2 j + x_3 k, \qquad\mbox{where}\qquad x_0,\,x_1,\,x_2,\,x_3\in\mathbbm{R},\qquad\mbox{and}\qquad i^2=j^2=k^2=-1.
\end{equation}
Quaternions are non-commutative because the imaginary units are anti-commutative and satisfy, by way of example, $ij=-ji$. 
In this article we adopt the symplectic notation for quaternions, where
\begin{equation}\label{i2}
q=z_0+z_1j,\qquad\mbox{so that}\qquad z_0=x_0+x_1i\qquad\textrm{and}\qquad z_1=x_2+x_3i.
\end{equation}
The quaternionic non-commutativity is algebraic and global, something very different from the local and geometric non-commutativity of the space-time coordinates in the formulation of Alain Connes \cite{Connes:2000ti}. An example of application of non-commutative geometry in physics with several references can be found in \cite{Abyaneh:2019cfk}. 
One could naively expect that $\mathbbm{H}$QM generalized quantum mechanics by introducing an additional complex degree of freedom, and consequently increasing the range of physical phenomena that can be described. However, the anti-commutativity constrains the theory, 
and the net balance of the introduction of quaternions in quantum mechanics is still an open question.

Let us then consider the quaternionic Schr\"odinger equation
\begin{equation}\label{w01}
\left[-\frac{\hbar^2}{2m}\nabla^2+V\right]\Psi=\hbar\frac{\partial\Psi}{\partial t}i.
\end{equation}
where $V=V(\bm x)$ is a quaternionic potential and $\Psi=\Psi(\bm x,\,t)$ is a quaternionic wave function. We point out that the
 the imaginary unit $\,i\,$ multiplies the right hand side of the time derivative of the wave function. 
Proceeding in analogy to $\mathbbm C$QM \cite{Schiff:1968qmq}, we consider a two particle scattering
problem and rewrite (\ref{w01}) as two equations. One equation describes the motion of the center of mass with energy $E_c$, and another equation describes the relative motion between the two particles with energy $E$. If $\,\mathcal{E}\,$ is the total energy of the system, the wave function is
\begin{equation}\label{w02}
\Psi\,=\,w\,u\,e^{-i\frac{\mathcal{E}}{\hbar}t}\qquad\mbox{and}\qquad\mathcal{E}=E_c+E,
\end{equation}
where $u=u(\bm x)$ and $w=w(\bm x)$ are quaternionic functions and a comprehensinve solution of the time-dependent equation can be found in 
\cite{Giardino:2017yke,Giardino:2017nqs}. Separating the functions, the time independent Schr\"odinger equation gives
\begin{equation}\label{w03}
\left[-\frac{\hbar^2}{2\mu}\nabla^2+V\right]u=Eu,\qquad\textrm{and}\qquad-\frac{\hbar^2}{2m}\nabla^2w=E_cw.
\end{equation}	
The center of mass behaves as a free particle of total mass
$\,m=m_1+m_2\,$ and wave function $w$, and the scattering is described by the wave function $u$, where $\,\mu\,$ is the reduced mass of the system. Our aim is to determine the cross section for a cylindrically symmetric
potential $\,V\,$ in the center of mass coordinate system, where $\,r\,$ is the distance between the two particles.
 Let us recall that in the complex case the general solution of  (\ref{w03}) is
\begin{equation}\label{w04}
u_\mathbbm{C}(r,\,\theta)=\sum_{\ell=0}^\infty\big(2\ell+1\big)\,R_\ell(r)(i)^\ell \,P_\ell(\cos\theta)
\end{equation}
where $\,P_\ell\,$ is a Legendre polynomial and $R_\ell$ depends on spherical Bessel functions \cite{Schiff:1968qmq}. Our aim is to propose a quaternionic solution that generalizes (\ref{w04}). Using the symplectic decomposition (\ref{i2}), we rewrite  (\ref{w04}) by decomposing $R_\ell$ as follows $\,R_\ell= R_\ell^{(0)}+R_\ell^{(1)}j,\,$ where $\, R_\ell^{(0)}\,$ and $\,R_\ell^{(1)}\,$ are complex functions. In the quantum complex case, the asymptotic behavior is expected to be
\begin{equation}\label{w05}
u_\mathbbm{C}(r\to\infty,\,\theta)\propto \; C_0 e^{ikz}\,+\,f(\theta)\frac{e^{ikr}}{r},
\end{equation} 
where $C_0$ is a complex constant, $f(\theta)$ is a complex function,  $k=\mu v/\hslash$,  and $v$ is a velocity ascribed to the reduced mass. The complex exponential has to be expanded in
Legendre polynomials in order to be compared with (\ref{w04}). 
In the quaternionic case, we propose a series using quaternionic coefficients and replace the complex function $f(\theta)$ with a quaternionic function $F(\theta)$, so that
\begin{equation}\label{w06}
	u(r\to\infty,\,\theta)= \; \frac{1}{kr}\sum_{\ell=0}^\infty\Lambda_\ell (2\ell+1)(i)^\ell\sin\left(kr-\frac{\ell\pi}{2}\right)
P_\ell(\cos\theta)\;+\;F(\theta)\frac{e^{ikr}}{r},
\end{equation} 
where $\,\Lambda_\ell\,$ are the unitary quaternionic coefficients
\begin{equation}\label{w07}
\Lambda_\ell=\cos\Theta_\ell\, e^{i\delta_\ell}\,+\,\sin\Theta_\ell\, e^{i\xi_\ell}\,j,\qquad\mbox{so that}\qquad \overline{\Lambda}_\ell\Lambda_\ell=1.
\end{equation}
The complex result is recovered for $\,\Lambda_\ell=1\,$ and $\,F(\theta)\to f(\theta).\,$
 The incident quaternionic particle is consequently a linear combination of quaternionic free particles that generalizes the first term of (\ref{w05}). We  obtain quaternionic scattering solutions from (\ref{w04}) using 
asymptotic quaterninic expansions for $R_\ell$ in  and imposing the equality with (\ref{w06}). Let us then assume the asymptotic expansion
\begin{equation}\label{w08}
R_\ell\propto \frac{1}{2kr}\Big[\,\Lambda_\ell\, e^{i\left(kr-\frac{\ell\pi}{2}\right)}\,-\,\overline{\Lambda}_\ell\, e^{-i\left(kr-\frac{\ell\pi}{2}\right)}\,\Big],
\end{equation}
and consequently (\ref{w04}) becomes
\begin{equation}\label{w09}
u(r,\,\theta)\,=\,\frac{1}{kr}\sum_{\ell=1}^\infty(2\ell+1)A_\ell\left[\,i\cos\Theta_\ell\sin\left(kr-\frac{\ell\pi}{2}+\delta_\ell\right)\,+\,\sin\Theta_\ell\,\cos\left(kr-\frac{\ell\pi}{2}\right)\,e^{i\xi_\ell} \,j\,\right](i)^\ell P_\ell(\cos\theta),
\end{equation}
where also $A_\ell$ is quaternionic and the complex case is recovered after imposing $\Theta_\ell=0$. Following \cite{Giardino:2020cee}, we interpret $\,\Theta_\ell\,$ a polarization angle that combines the pure complex and the pure quaternionic components of the quaternion basis elements, which are orthogonal \cite{Giardino:2018lem,Giardino:2018rhs}.
The asymptotic expressions (\ref{w06}) and (\ref{w09}) must be compatible, and the equality of both of the expressions give
\begin{equation}
A_\ell=-\Lambda_\ell i \Lambda_\ell,
\end{equation}
and consequently
\begin{equation}\label{w10}
F(\theta)=\frac{1}{2k}\sum_{\ell=0}^\infty (2\ell+1)\,\Lambda_\ell\,i\,\Big(1-\Lambda^2_\ell\Big)P_\ell(\cos\theta).
\end{equation}
We can now study several features of this wave function. 
The cross-section is defined by $\,\sigma(\theta)=|F(\theta)|^2\,$ and  the total cross section $\,\sigma\,$ is obtained 
from the integral over the scattering angle, so that $\sigma=\int\sigma(\theta)d\Omega.\,$ Using the orthogonality conditions for Legendre polynomials, we get
\begin{equation}\label{w11}
\sigma=\frac{4\pi}{k^2}\sum_{\ell=0}^\infty\big(2\ell+1)\Big(\sin^2\delta_\ell\cos^2\Theta_\ell+\sin^2\Theta_\ell\Big).
\end{equation}
In agreement with the complex case, a finite cross section requires evanescent phase and polarization angles, so that $\,\delta_\ell,\,\Theta_\ell\to 0,\,$ 
when $\ell\to\infty.\,$ Additionally, we recover the complex cross section in the  $\Theta_\ell=0\,$ complex limit, leading to a complete agreement between the complex and quaternionic cases. Conversely, the $\,\delta_\ell=0\,$ limit generates a new quaternionic solution whose cross section reproduces the
complex result. However, the important physical interpretation is that the $\mathbbm{H}$QM predicts a larger cross section when compared to
$\mathbbm{C}$QM, and thus we expect a more intense scattering in the quaternionic case

The phase shifts $\delta_\ell$ and the polarization angles $\Theta_\ell$ also ascertain the difference between the asymptotic form of the wave function and the exact solution of the Schr\"odinger equation (\ref{w03}). The scalar potential 
$\,V\,$ plays an important 
role defining a radial distance $\,r=a\,$ that determines  the region of the space where $\,V(r>a)\approx 0\,$ and the asymptotic approximation holds. By analogy to the complex case, we estimate the phase shifts imposing that the continuity of the radial wave function $\,R_\ell\,$ at $\,r=a\,$  depends on a quaternionic  constant $\,\Gamma_\ell,\,$ so that
\begin{equation}\label{w12}
\frac{1}{R_\ell}\frac{dR_\ell}{dr}\Big|_{r=a}=\Gamma_\ell\qquad\qquad\mbox{where}\qquad\qquad \Gamma_\ell=\Gamma_\ell^{(0)}+\Gamma_\ell^{(1)}\,j,\qquad
\mbox{and}\qquad \Gamma_\ell^{(0)},\,\Gamma_\ell^{(1)}\in\mathbbm{C}.
\end{equation}
Recalling the wave function (\ref{w09}) as the asymptotic limit of 
\begin{equation}\label{w13}
R_\ell(r)=A_\ell\left[\,i\cos\Theta_\ell\Big(\cos\delta_\ell\, j_\ell(kr)\,-\,\sin\delta_\ell\, y_\ell(kr)\Big)\,-\,
\sin\Theta_\ell\, e^{i\xi_\ell}\,y_\ell(kr)\,j\,\right],
\end{equation}
and using $j_\ell,\,j'_\ell,\,y_\ell\,$ and $\,y'_\ell\,$ for spherical Bessel functions and their derivatives at $r=a$, we eventually obtain
\begin{equation}\label{w14}
\Gamma_\ell^{(0)}\,=\,k\,\frac{y_\ell y'_\ell\tan^2\Theta_\ell\,+\,\Big(y_\ell'\sin\delta_\ell+j'_\ell\cos\delta_\ell\Big)\Big(j_\ell\cos\delta_\ell-y_\ell\sin\delta_\ell\Big)}
{y_\ell^2\tan^2\Theta_\ell\,-\,\Big(y_\ell\sin\delta_\ell-j_\ell\cos\delta_\ell\Big)^2}.
\end{equation}
We remember that (\ref{w14}) recovers the complex result for $\Theta_\ell=0$, as expected. Furthermore, we also obtain
\begin{equation}\label{w15}
\Gamma^{(1)}_\ell\,=\,\tan\Theta_\ell\,\frac{\Gamma_\ell^{(0)}y_\ell-k y'_\ell}{\cos\delta_\ell\, j_\ell\,-\,\sin\delta_\ell\, y_\ell}\,
e^{i\left(\xi_\ell+\frac{\pi}{2}\right)},
\end{equation}
which accordingly disappears for $\Theta_\ell=0$ as expected. 
We finally observe that $\,\Gamma_\ell^{(0)}\,$ is a real quantity, in further agreement with the complex scattering, while $\,\Gamma_\ell^{(1)}\,$ is indeed complex. On the other hand, 
we also have that
\begin{equation}\label{w16}
 \tan\delta_\ell\,=\,-\,\frac{\left|\Gamma_\ell^{(1)}\right|^2 y_\ell \,j_\ell-\Big(\Gamma_0j_\ell-k\,j'_\ell\Big)\Big(\Gamma_0y_\ell-k\,y'_\ell\Big)}
{\left|\Gamma_\ell^{(1)}\right|^2 y^2_\ell-\Big(\Gamma_0 y_\ell-k\,y'_\ell\Big)^2}.
\end{equation}
The $\mathbbm{C}$QM result is accordingly recovered for $|\Gamma_\ell^{(1)}|=0$, as  desired. This analysis permit us to be sure that
the asymptotic limit can be obtained continuously, without the risk of singular points and exotic unphysical situations.

\section{\sc The rigid sphere\label{E}}
In this section, we examine the simplest example for the quaternionic scattering: the
 perfectly spherical rigid potential. In spherical coordinates,
\begin{equation}\label{e01}
V(r)=\left\{
\begin{array}{ll}
\infty &\qquad\mbox{for}\qquad r < R\\
0      &\qquad\mbox{for}\qquad r\geq R.
\end{array}
\right.
\end{equation}
This potential is slightly different from the usual rigid spherical potential, because the point
$r=R$ belongs to the domain of the wave function, a strategy that has already been used in the quaternionic square well potential \cite{Giardino:2020cee}, and that permits a finite value for the wave function 
at the surface of the sphere, while the wave function is set identically zero at this point in the  complex solution \cite{Schiff:1968qmq}. In order to get agreement with 
the complex case, we adopt a boundary condition where only the pure complex component of the wave function (\ref{w13}) vanishes at $r=R$. In this situation, 
\begin{equation}\label{e02}
\Gamma^{(0)}_\ell=k\frac{y'_\ell(kR)}{y(kR)}\qquad\mbox{and}\qquad\tan\delta_\ell=\frac{j_\ell(kR)}{y_\ell(kR)}.
\end{equation}
The phase angle $\,\delta_\ell\,$ is absolutely identical to the complex case. Adopting the low energy limit, where $\,kR\ll 1,\,$ then
\begin{equation}\label{e03}
\tan\delta_\ell=-\frac{(kR)^{2\ell+1}}{\big(2\ell+1\big)!!\big(2\ell-1\big)!!}.
\end{equation}
Additionally, the finiteness of the pure quaternionic component is obtained from 
\begin{equation}\label{e04}
\sin\Theta_\ell=\frac{1}{y_\ell(kR)}\approx -\,\frac{(kR)^{\ell+1}}{\big(2\ell-1\big)!!}.
\end{equation}
Consequently, both of the phase shifts rapidly go to zero with increasing $\ell$. Considering the $\ell=0$ component of
the total cross section (\ref{w11}) and using $\,\sin\delta_0=-\sin\Theta_0=kR\,$, we obtain 
\begin{equation}\label{e05}
\sigma\approx 8\pi R^2\left(1-\frac{1}{2}k^2R^2\right),
\end{equation}
and the result is almost twice the $\mathbbm{C}QM$ result. In high  energy limit, where $kR\gg 1$,  we use (\ref{e02}) to obtain
\begin{equation}\label{e06}
\sigma=\frac{4\pi}{k^2}\sum_{\ell=0}^\infty\big(2\ell+1)\frac{\,j^2_\ell(kR)\,+\,\sin^2\Theta_\ell\, y^2_\ell(kR)\,}
{j^2_\ell(kR)+y^2_\ell(kR)}\,,
\end{equation}
a result that recovers the complex result for $\Theta_\ell=0$, as desirable.

\section{\sc The optical theorem\label{O}}
In a region without sources or sinks of probability, we expect that 
\begin{equation}\label{o01}
\oint_\mathcal{S} \bm{J\cdot}d\bm\sigma=0,
\end{equation}
where $\,\bm{J}\,$ is the probability current density and $\,\mathcal{S}\,$ is a closed surface. We remember 
the probability current density in real Hilbert space $\mathbbm{H}$QM \cite{Giardino:2018lem,Giardino:2018rhs} to be defined as
\begin{equation}\label{o02}
\bm J=\frac{1}{2m}\Big[\,\overline\Psi\,\bm\Pi\Psi+\left(\overline{\bm\Pi\Psi}\right)\Psi\,\Big],\qquad\qquad\mbox{where}\qquad\qquad
\bm\Pi\Psi=-\,\hbar\bm\nabla\Psi\,i.
\end{equation}
Defining (\ref{w06}) to be the wave function and the sphere to be the closed surface, from (\ref{o01}) and $\,\sigma=\int |F|^2d\Omega\,$ we obtain
\begin{equation}\label{o03}
\sigma=-\frac{1}{2k}\,\oint\,r^2\left[\,\bar I\frac{\partial I}{\partial r}i\,+\,\frac{1}{r}\left(e^{-ikr}\,\bar F\,\frac{\partial I}{\partial r}i-k\,\bar I\, F\,e^{ikr}\right)-
\frac{1}{r^2}\,\bar I\,F\,i e^{ikr}\,+\,\textrm{c.c.}\,\right]d\Omega
\end{equation}
where $\,I=I(r,\,\theta)\,$ corresponds to the incident particle element of the wave function (\ref{w06}), namely
\begin{equation}\label{o04}
I(r,\theta)\,=\,\frac{1}{kr}\sum_{\ell=0}^\infty\Lambda_\ell (2\ell+1)(i)^\ell\sin\left(kr-\frac{\ell\pi}{2}\right)
P_\ell(\cos\theta).
\end{equation}
Using the orthogonality of Legendre polynomials, 
\begin{equation}\label{o05}
\int_0^\pi P_\ell(\cos\theta)P_m(\cos\theta)\sin\theta d\theta=\frac{2}{2\ell+1}\delta_{\ell m}
\end{equation}
we obtain that
\begin{equation}\label{o06}
r^2\,\oint\,\left(\,\bar I\frac{\partial I}{\partial r}i\,-\,i\frac{\partial\bar I}{\partial r} I\right)d\Omega=0.
\end{equation}
We also observe that the term $|F|^2$ cancels because of the complex conjugation. Furthermore, using
\begin{equation}\label{o07}
\int_0^\pi P_\ell(\cos\theta)\sin\theta d\theta=\delta_{\ell 0}
\end{equation}
and integration by parts, we only keep the $\ell=0\,$ terms in (\ref{o04}) because t
The trigonometric terms highly oscillate
in the limit of large $\,r\,$ and does not contribute to the result. Thus, 
\begin{eqnarray}\label{o08}
\nonumber\sigma&=&\frac{1}{2}\,\oint\,\left(\,iF-\bar F i\,\right)d\Omega\\
\nonumber &&\\
 &=&2\pi\int_0^\pi\frak{Im}\big[\,iF\,\big]\sin\theta d\theta.
\end{eqnarray}
The above result is  similar to the complex case, where the cross section depends linearly to the imaginary component of the complex amplitude $\,|f(\theta)|.\,$ The quaternionic cross section agrees to the continuity equation for quaternionic scalar potentials \cite{Giardino:2018lem,Giardino:2018rhs}, where the imaginary unit $\,i,\,$ associated to momentum and energy operators, is physically more relevant than the purely quaternionic imaginary units. An integrated expression for the optial theorem is an important direction for future research.

\section{\sc Conclusion \label{C}      }

In this article we presented a quaternionic version for the elastic scattering, and we obtained a sound agreement with the complex results. This concordance reinforces the hypothesis for a viable quaternionic quantum mechanics in real Hilbert space. Directions for future research 
include the interesting phenomena of inelastic scattering. From a theoretical standpoint, real Hilbert space $\mathbbm H$QM seems mature to be applied in the exciting area of non-hermitian complex quantum mechanics, that much developed in recent times and was never studied in quaternionic terms.

%
%
%
%

\bibliographystyle{unsrt} 
\bibliography{bib_qscatt}
\end{document}